# Generation of electromagnetic radiation by a heat flow in He II

A.S. Rybalko, S.P. Rubets, E.Ya. Rudavskii, V.A. Tikhiy, R. Golovachenko[*], V.N. Derkach[*], S.I. Tarapov[*]

B.I.Verkin Institute for Low Temperature Physics and Engineering, National Academy of Sciences of Ukraine, 47, Lenin Ave., Kharkov, 61103.

[*] Institute for Radiophysics and Electronics, National Academy of Sciences of Ukraine, 12, St. Proskury, Kharkov, 61085.

rybalko@ilt.kharkov.ua

Abstract

A new effect of electromagnetic radiation generated by heat flow in superfluid helium has been detected experimentally. The generating heat flow was produced using hydrodynamic thermal guns. Electromagnetic radiation was registered with a dielectric disk resonator in the interval 1.4K –2.17K. The power of the thermal gun heater varied within Q up to $10^{-3}$ W. A distinct signal of electromagnetic radiation was registered when the thermal gun was on and there was no signal at Q=0. The frequency of the generated electromagnetic radiation was measured with a Fabry-Perrot interferometer. It was about 180·GHz at T=1.4K and decreased with a rising temperature practically following the temperature dependence of a roton gap. The detected signal was investigated as a function of the heat flow temperature, direction and power Q. The results obtained show that He II behaves as a two – level system in which the difference between the levels is equal to the roton energy.

## 1. Introduction

A number of new interesting effects connected with the interaction between superfluid helium (moving or at rest) and an electromagnetic wave have been detected [1-4] recently in a series of microwave experiments on He II in Kharkov. In particular, resonance absorption of the electromagnetic wave at a frequency corresponding to the minimal roton energy was observed in addition to ordinary dielectric losses. The observed spectral line consists of two parts – a very narrow line of resonance absorption and broad side wings ("background") [4]. The width of the resonance in the wings is similar to the roton line width in neutron scattering experiments [5,6]. A resonance line was unobservable in these [5,6] experiments, presumably because of inadequate resolution.

The microwave experiments [1-4] were made employing a dielectric disk resonator immersed into liquid helium. An external generator was used to produce "whispering gallery" waves that propagated around the cylindrical surface of the resonator. On passing through the resonator, the wave arrives at the detector that was at room temperature. The resonance line was observable when its frequency coincided with those of roton energy and of the whispering gallery modes. Circular flows of the liquid were also initiated artificially around the resonator using two "thermal guns". These were heat – insulated flasks with a heater and a thermometer inside. They were connected to a helium bath through a nozzle. Using this heat-flow technique (developed and tested by Kapitza [7]), we were able to generate counter-flow of the normal and superfluid components in He II. The velocities of the counter-flows were controlled by varying the power $Q$ onto the heater of the thermal gun. The circular flows around the resonator occurred [1-4] when the heat jet was tangential to the cylindrical surface of the dielectric. The appearance and change of flow rates could be registered by amplitude change of the signal detected. Consequently, one registers, in experiment, the resonance absorption of electromagnetic waves and also circular flows of superfliud liquid are created.

The character of the resonance line depends drastically on the heat applied: as the power Q increases, resonance absorption of the electromagnetic wave transforms into induced wave radiation [2], but the "background" remains essentially unaltered. The resonance line in the microwave spectrum appearing against the wing background, its splitting in the external electric field and the absorption – radiation processes [4] suggest that He II may be considered as a two-level system with the level difference equal to the roton energy. With the gun off, the energetically preferable lower level has excess occupancy. As the liquid starts to experience a microwave, the microwave photons are observed at the roton (resonance) frequency.

The mechanism of the liquid helium – electromagnetic field interaction in the roton – frequency region is not clear yet. Neither is evident how the production of a single roton by a microwave photon is complied with the law of momentum conservation.

## 2. The idea of the experiment

Since in our experiments He II behaves as a two-level model, it is interesting to examine generation of microwave photons more accurately using no external microwave generator in the measuring system. In this case, the electromagnetic wave, if it appears, can be related to the properties of superfluid helium, which is an additional reason for the presence of a narrow line in the spectrum of helium. The goal of this study was to search for this effect when the processes proceeding in He II act as a quantum generator of microwave oscillations.

It is known that excess occupancy of the excited level is a necessary condition for a two-level system to generate electromagnetic radiation. There are two ways of exciting the upper level – thermal or electromagnetic. The latter was used in [1] and it would be therefore interesting to test the other one. For this purpose the external generator was removed from the measuring circuit and thermal guns 1 (see Fig. 1) with heaters and thermometers inside were used instead (as in our previous experiments). According to the Boltzmann distribution, the excess population of the excited state was determined by the temperature in the gun flask and could be up to $3 \cdot 10^{-3}$ K higher than in the outer bath (linear regime). It is also essential that the regions of particle excitation and relaxation should be spaced apart (as in a hydrogen maser [8]) where excited atoms are carried by a particle beam to a generator and relax there to the lower level). According to two-liquid hydrodynamics, the gun nozzle emits a weakly spraying normal component jet with highly populated higher-lying energy levels. The temperature in heat flow is some higher than that of bath. As a result, the particle density is higher in upper energy level. The system relaxes in outer He II bath on some distance from heat gun flask and the transition in ground state is accompanied by spontaneous photon emission. To make possible register the generated wave, the process should occur near dielectric disk resonator.

Electromagnetic generation requires that the energy losses in the receiving resonator should be lower than the induced radiation energy. The requirement was met using a quartz resonator generating high–Q whispering gallery modes ($Q \approx 10^6$ when the resonator was immersed in superfluid helium).

### 3. Experimental technique

The basic diagram of the measuring system is illustrated in Fig. 1. Unlike the previous experiments [1-4], it has no high-frequency generator, which was replaced with detector as before. It consists of two parts: a system generating directed heat flows in He II and an electrodynamic block for measuring the electromagnetic radiation induced by heat flows. Two thermal guns 1 and 1a with heaters and thermometers inside generate directed heat flows in He II. The gun nozzles are oriented so that the heat flow was tangential to the cylindrical surface of disk resonator 2. Unlike the previous experiments, waveguides 3 both operated as receiving antennas.

Modulators 4 placed after the receiving antennas (at room temperature) served as switches and could interrupt electromagnetic radiation. The modulators were controlled by low-frequency (below 1 kHz) square – wave generator 5. The signal then arrived of standart microwave detecting heads 6 (3A129 Shottky diode) sending the signal to amplifiers 7 with synchronous detectors. With this receiving system the measurement sensitivity was improved to $10^{-11}$ W/Hz$^{0.5}$, as estimated from the noise track. Two symmetrically arranged antennas (see Fig. 1) were pasted into waveguides 3. Each of them registered a whispering gallery wave in a certain direction. The symmetrical system with two detectors and modulators enabled us to register the whispering gallery waves generated by a directed heat flow and follow straightway the correlation between the directions of the heat flow and the electromagnetic wave. The symmetry was only disturbed when a wavemeter is connected to the measuring system behind one of the modulators (4a) to measure the frequency of electromagnetic radiation.

The radiation wavelength was measured with a standard DL-2116 wavemeter based on Fabry-Perrot resonator 8: its partially transparent diffraction grate served as entry mirror 9 and its bell concentrated radiation onto detector 6a. Piston 11 of the Fabry – Perrot resonator was moved using a special motor (the rest of components are evident in the diagram). The signal of electromagnetic radiation ran from the modulator to the mirror and then its reflected part was registered by detector 6a. The other part of the signal was sent to interferometer 8 by moving piston 11. When an integer number of half- waves could fit within the distance in the interferometer, detector 6a registered the lowest signal. The received signals were registered, and the electromagnetic radiation wavelength was estimated from the distance between the neighboring minima. The temperature interval was T>1.4K. The power onto the heater of the thermal gun could be up to $Q \approx 5 \cdot 10^{-3}$W, which corresponds to the heatflow in the nozzle $Q/S \approx 10$W/cm$^2$. 5 is the nozzle section.

## 4. Results and discussion.

The fact of generation of electromagnetic radiation in superfluid helium by a direct heat flow is evident in Fig. 2. Detector 6a registers a distinct signal (Fig. 2b) at the moment a when the power Q/S=7W/cm$^2$ is fed to thermal gun 1a (Fig. 2a). There is no signal at Q=0. The amplitude of the signal responds readily and then switched on again. This is unambiguous evidence for well-defined correlation between switching on a heat flow and the signal at the detector.
The readings of the thermometer inside the gun flask show that the steady – state temperature is typically achieved there in 2-4 seconds, and the characteristic time of signal relaxation at the detector is several tens of seconds due to the high Q-factor of the resonance line.
When gun 1a was on, the signal was registered only by detector 6a. The other detector was insensitive to switching on and off gun 1a. On switching on the other gun, detector 6a remained blind and the signal was registered by other detector. with two guns on, the signal was received by both detectors, which can reasonably be interpreted as electromagnetic radiation propagating through waveguides from superfluid helium.

The heat flow in He II usually excites the counter-flow of the normal and superfluid components. The normal component velocity is oriented along heat flow. The fact that, under switching on of one among heat gun, the generation is registered by only one detector, argues that running (not standing) whispering gallery wave is excited in the resonator. The direction of its penetration can be changed if we switch one or other thermal gun.

In experiment, at maximum heat power Q of $4 \times 10^{-3}$ W on the gun heater (the specific power in the gun flask is up to ~10W/cm$^2$), the overheat of He II inside the gun was ~$3 \cdot 10^{-3}$ K. As it was demonstrated in Ref. 9, the increase of Q in He II flow from the gun leads to a transition from laminar to turbulent regime. In present work this effect was not studied. However the results obtained support the supposition that generation effect is observed in both laminar and turbulent flow.

It is also evident that the signal amplitude is dependent on the temperature of He II. In Fig. 3 this dependence is illustrated for a constant superheating temperature inside the gun flask $\Delta T = 3 \times 10^{-3} K$, which is higher than the bath temperature. The generation of electromagnetic radiation is most efficient at T=1.8-2.0K but the reason for this is not evident. The power of the detected signal was typically $\approx 10^{-6}$W, or $\approx 10^{-5}$W taking into account the attenuation in the waveguide track. The radiation power was measured in both of the symmetrical channels (the wavemeter was removed from the measuring system). The dependence of the signal amplitude upon the power onto the thermal gun heater is illustrated in Fig. 4 for several temperatures. One can see the power of irradiation increases with helium overheating temperature in chamber even at rather high specific power of heat flow.

The next important step was to measure the frequency of the generated electromagnetic radiation using a Fabry - Perrot interferometer. The typical dependence of the signal amplitude on the position of the piston in the Fabry-Perrot resonator is shown in Fig. 5. The movement of the piston was calibrated in wavelength units. The distance traveled by the piston is 9.96 mm between the neighboring minima. The signal is minimal when the piston passes through the positious corresponding to a half wave integer. The distance between the neighboring modes is 1.66mm (Fig. 5), which corresponds to the radiation frequency $\approx 1.8 \cdot 10^{11}$ Hz at T=1.4K.

Note that the frequency of the generated electromagnetic radiation practically coincides with the roton frequency at this temperature. It is also interesting that the excited electromagnetic wave is always opposite to the heat flow, i.e. its direction coincides with the velocity vector of the superfluid component. The fact needs to be explained.

We also measured the power of electromagnetic radiation at the detectors. The obtained value $W_{em} \approx 10^{-5}$W agrees in order of magnitude with the power of the mechanical motion of the normal component in the heat flow (estimated from [9-11]). This means that the momenta of both liquid flow and photons of electromagnetic field are the same though oriented in opposed directions. As is known the particle obtains the output momentum *hf/c*, if a photon emitted. Here *hf = E* is photon energy and *c* is light velocity. Unfortunately it is not clear at the moment what is the reason for emission: quasiparticle, atom or elementary particles forming atom. If one supposes that output momentum is absorbed by atom, the atom obtains the velocity ~1mm/s in addition to thermal velocity. This, in conditions of some ordering in the liquid, could lead to arising a drift velocity even under thermal equilibrium. Possibly, the output momentum may be a source of mechanical undamped motion in He II. This problem should be studied additionally.

Our experiments show that the energy emitted from the gun heater ($10^{-3}$W) permits the induced heat flow to generate radiation of $\approx 10^{-5}$W at a fixed frequency. This means that in our system the transformation coefficient of the gun thermal energy to electromagnetic radiation is about 1%.

## 5. Conclusion

The described series of experiments have led us observe for the first time generation of UHF electromagnetic radiation in He II applying a heat flow. The genesis of this radiation can be explained within the two-level model for some subsystem of superfluid helium. The heat flow excited by the heater of a thermal gun increases the occupancy of the upper excited state depending on the He II temperature in the heat flow region. As soon as the heat flow comes in contact with the cylindrical surface of the resonator, the retarding processes in the normal component relax the excited state and provoke emission of photons. The atoms can emit only in the direction opposite to the flow. When the frequency of the one of the high-Q whispering gallery modes generated in the dielectric disk resonator coincides with the radiation frequency, the oscillations become enhanced and synchronized.

According to the two–liquid model, below the $\lambda$-point the part of the liquid defined as a superfluid component which microscopic nature is not clear yet. The superfluid momentum was introduced phenomenologically basing on experimental observation of circular undamped flow and non-classical rotation inertia [13] is of quantum-mechanical origin and is a characteristic of macroscopic equilibrium energy state. The results of present experiments give grounds to suppose that output momentum obtained by individual particle under photon emission with roton energy is absorbed by the system of all atoms in analogy with Mossbauer effect where the crystalline grate absorbs the output momentum. These suppositions do not contradict to two-fluid hydrodynamics. We also trust that the methodics developed in present work could be perspective in a study of electrodynamic properties of He II.

The authors are indebted to A. Andreev, V. Grigoriev, V. Loktev, V. Maidanov, E. Pashitsky, S. Ryabchenko, M. Tomchenko, S. Shevchenko and other participants of the International Conference on Low Temperature Physics LT25(Amsterdam, 2008) and the Ultralow temperature Symposium ULT2008 (London, 2008) for their interest shown to this research and for helpful discussions.

This study was supported by CRDF (Project 2853) and STCU (Project 3718) grants.

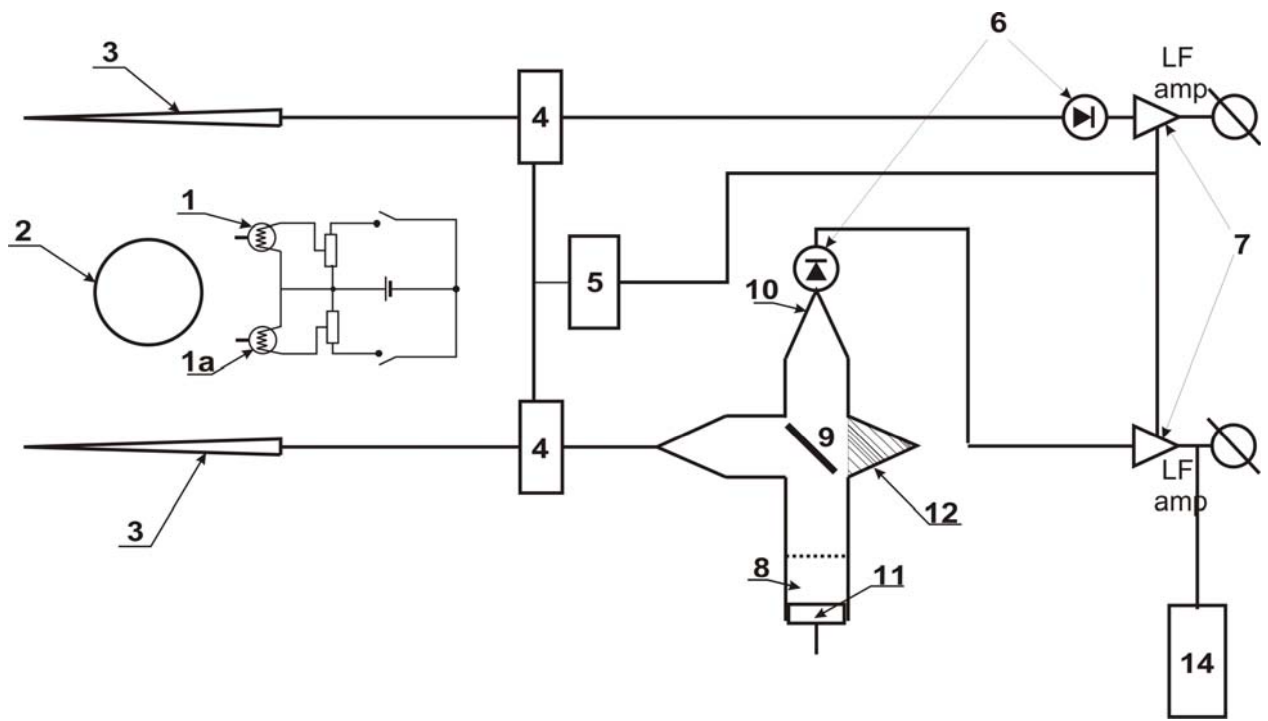

Rice 1. The basic scheme of measuring system for registration of the electromagnetic radiation induced by a thermal stream in II. 1 – thermal guns, 2 – a body of the dielectric disk resonator, 3 – aerials with wave guides, 4 – modulators, 5 – the generator of rectangular impulses of low frequency, 6 – detectors of the MICROWAVE of radiation, 7 amplifiers of low frequency, 8 – the resonator Fabry-Perot, 9 - an entrance mirror in the form of partially transparent дифракционной lattices, 10 – a bell, 11 – the piston, 12 – an absorber.

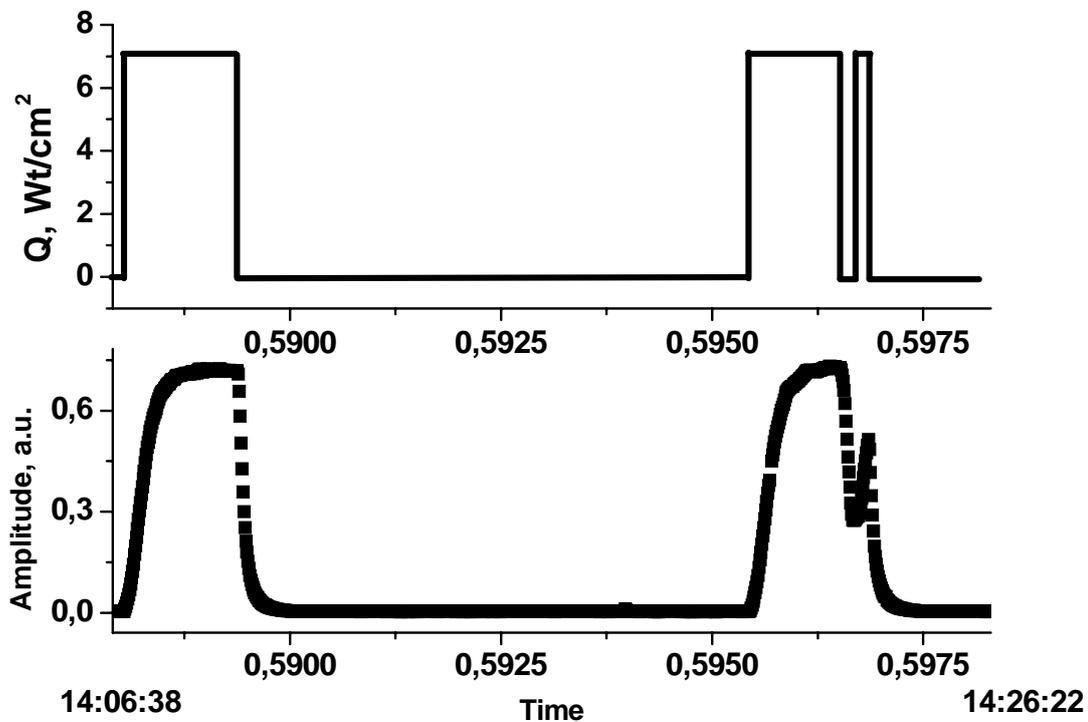

Fig. 2. Dependence on time of the capacity submitted on a heater of a thermal gun (a) and corresponding amplitude of the signal detected by the MICROWAVE by the detector (b).

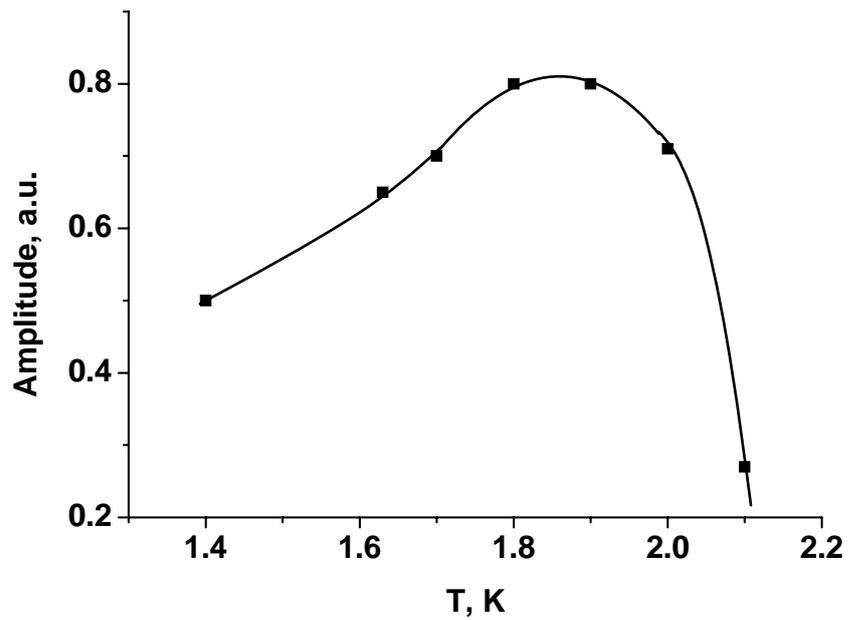

Fig. 3. Temperature dependence of amplitude of a detected signal at constant value of temperature of an overheat of interiors of a flask of a gun concerning a bath $\Delta T=3 \times 10^{-3}$ K.

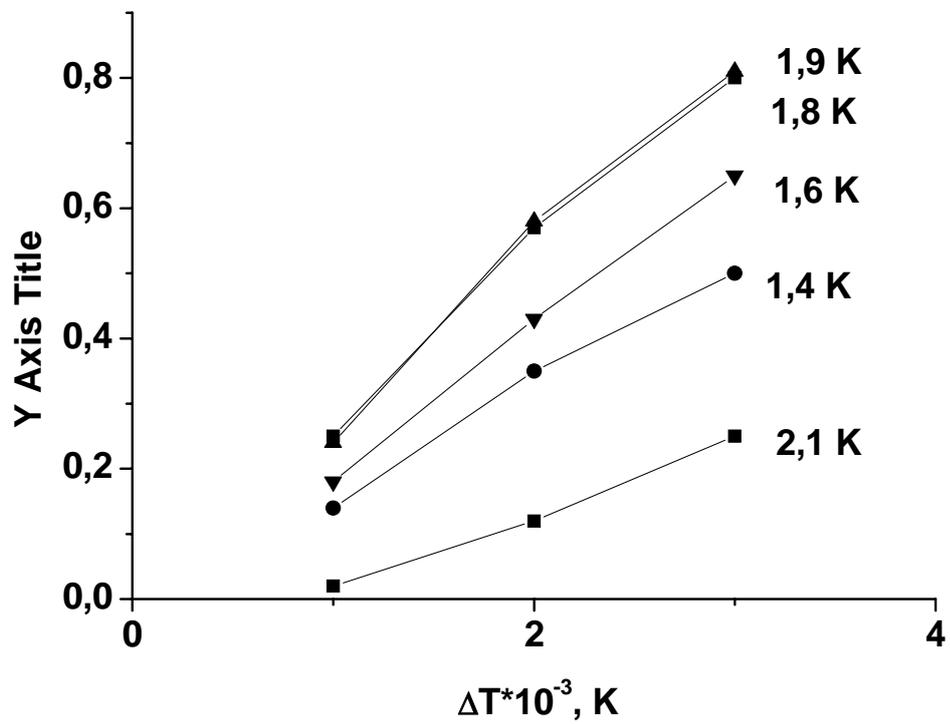

Fig. 4. Dependence of amplitude of a detected signal on the capacity submitted on a heater of a thermal gun at various temperatures: 1 – T-1,4 K, 2 – T-1,6 K, 3 – 1,78 K, 4 – 1,9 K, 5 – 2,1 K.

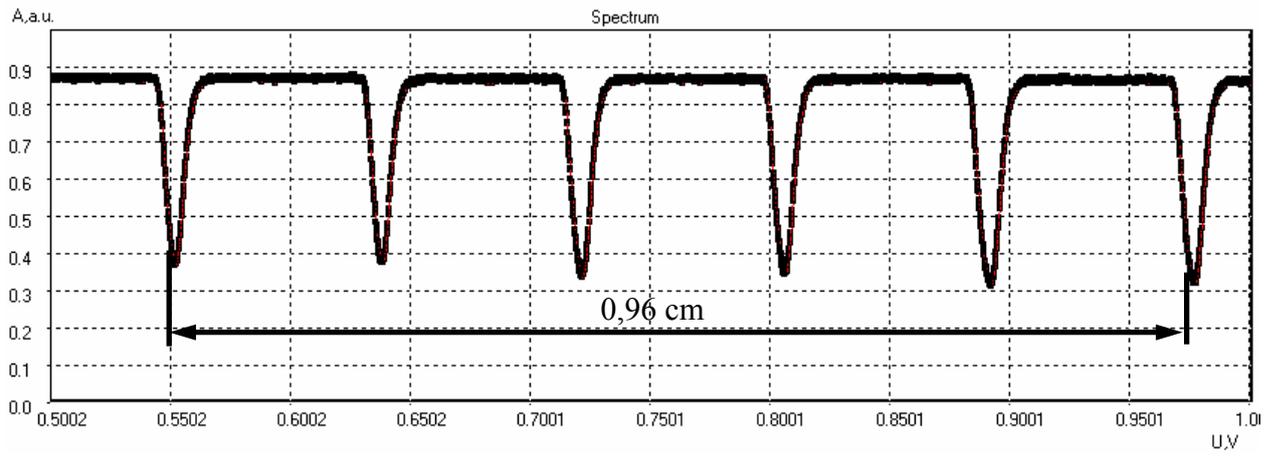

Fig. 5. Dependence of the signal amplitude on the position of the piston in the Fabry – Perrot resonator (temperature 1.4K).